\documentclass[a4paper,fleqn,usenatbib,twocolumn]{mnras}
\setcitestyle{}

\usepackage{newtxtext,newtxmath}

\usepackage[T1]{fontenc}
\usepackage{ae,aecompl}
\def\be{\begin{equation}}
\def\ee{\end{equation}}
\def\bea{\begin{eqnarray}}
\def\eea{\end{eqnarray}}
\newcommand{\bes}{\begin{subequations}}
\newcommand{\ees}{\end{subequations}}
\def\comment#1{}

\usepackage{color}

\usepackage{graphicx}	
\usepackage{amsmath}	

\usepackage{amssymb}	
\title{Gravitational-wave bursts from spin-precessing black holes in binary systems}

\author[C. Zhang, W.-B. Han, S.-C. Yang ]{
Chen Zhang$^{1}$, Wen-Biao Han$^{1,2,3,4}$\thanks{Corresponding author: wbhan@shao.ac.cn} , Shu-Cheng Yang$^{1}$\\
 $^1$Shanghai Astronomical Observatory, Shanghai, 200030, China\\
 $^2$School of Fundamental Physics and Mathematical Sciences, Hangzhou Institute for Advanced Study, UCAS, Hangzhou 310024, China \\
 $^3$School of Astronomy and Space Science, University of Chinese Academy of Sciences, Beijing 100049, China\\
 $^4$Shanghai Frontiers Science Center for  Gravitational Wave Detection, 800 Dongchuan Road, Shanghai 200240, China
 }

\date{Accepted XXX. Received YYY; in original form ZZZ}

\pubyear{2022}
\allowdisplaybreaks
\begin{document}
\label{firstpage}
\pagerange{\pageref{firstpage}--\pageref{lastpage}}
\maketitle

\begin{abstract}
Gravitational waves from precessing binary black holes exhibit new features that are absent in non-precessionary systems. All current waveform models take into account only the modulation of the signal due to precession. In this letter, we find that this effect has its own signature, by gravitational emission of a short and transient signal, or burst. The frequency of the burst is comparable to that of the late stage of the inspiral. We show that under certain conditions, this signal is strong enough to be detected by Advanced LIGO. For third-generation detectors like the Einstein telescope, the calculated signal-to-noise ratio can reach higher values.  Measurements of precession would provide valuable insights into the intrinsic structure of black holes, and therefore into astrophysical binary formation mechanisms.
\end{abstract}

\section{introduction}

All the 90 GW events released by the Advanced LIGO-Virgo-KAGRA collaboration are compact binary coalescences(CBCs) \citep{lvkGWTC3}. As the two compact objects inspiral towards merger due to the emission of gravitational waves(GWs), if one or both bodies are rapidly rotating, the general relativistic spin-orbit and spin-spin couplings(i.e., the ``dragging of inertial frames" by the bodies' spins) cause the orbital plane and the spins to precess about the direction of the total angular momentum \citep{apostolatos1994spin}. Thus spin-precession is an important phenomenological feature which relates to both general relativistic dynamics and astrophysical binary formation scenarios.

During the inspiral stage, gravitational waves are radiated from the time-varying quadrupole due to the orbital motion of two masses $m_1$ and $m_2$, after that, the merger-ringdown waveforms are produced by the perturbation of the dynamically forming black holes (BHs). The inspiral process can be clearly described by the post-Newtonian (PN) approximation for the quasi-circular orbits. A few of these events have been considered to have spin 
precessions, for example, the event GW190521 \citep{abbott2020gw190521,estelles2022detailed}, which has an effective precession spin $\chi_p$ as large as 0.67, shows evidence for spin-induced orbital precession.
In addition, GW190412, GW190512{\_}180714, GW190521, and GW190814 signals also provide sufficient support for spin precession\citep{gerosa2021generalized}. However, current waveform models only take into account the signal modulation induced by the spin precession on the inspiral waveform \citep{Pan2014PhRvD..89h4006P,hannam2014simple,Babak2017PhRvD..95b4010B,Khan2019PhRvD.100b4059K,Varma2019PhRvR...1c3015V}.

In this Letter, we propose that the spin-precession of a black hole itself also can produce detectable GWs. This is due to the variation of quadrupole from the precessing Kerr black hole. It is well known that the Kerr black hole has a quadrupole moment $Q = -M^3 \chi^2 $, for an isolated Kerr black hole, the quadrupole is invariant and can not radiate GWs because of axisymmetric spacetime. However, during the inspiral of precessing binary black holes, the quadrupole moment of black hole itself will change with time and radiate GWs. In this Letter, we calculate this precessing GWs and demonstrate the properties of this brand new signal, especially discuss the detectability.

Just a reminder, this precessing GWs  are different from the previous researches on the precessing binaries which only take into account the modulation induced by the spin precession on the inspiral waveform \citep{Pan2014PhRvD..89h4006P,hannam2014simple,Babak2017PhRvD..95b4010B,Khan2019PhRvD.100b4059K,Varma2019PhRvR...1c3015V}. As a consequence of the no-hair theorem---the claim that mass and spin are the only two properties needed to describe black holes in general relativity\citep{vishveshwara1970stability}, the quadrupole momentum should be determined entirely in terms of the mass $M$ and the dimensionless spin magnitude $\chi$ of the black hole. Present researches provide tests of the no-hair theorem at the $ \sim 10\%$ level by analyzing ringdown data from the Advanced LIGO - Virgo detection \citep{isi2019testing,ota2020overtones}. Considering the ringdown signals just reflect the properties at light-ring \citep{Cardoso2017NA}, the precessing waves are directly determined by the mass, spin, and quadrupole of the black hole, then this kind of signals could be a unique tool to test the no-hair theorem.

\section{Waveform emitted by black hole's precession}

There are two main channels for the formation of compact binaries: common envelope evolution\citep{fragione2021black,mandel2022rates,belczynski2022black} and dynamical capture\citep{vitale2017use,fragione2022repeated,fragione2020origin}. The latter one will induce spins that are randomly oriented for binaries. Precessing spins are a key prediction of binary black holes formed in dense clusters, but might also be present in the case of sources formed in isolation because of, for example, supernova kicks\citep{kalogera2000spin,mandel2010compact,gerosa2013resonant,rodriguez2016illuminating,gerosa2018spin,steinle2021pathways,callister2021state,fragione2021impact}. This spin precession could occur during the merger of binary black holes, if the spins are not aligned with the orbital angular momentum, a so-called geodetic precession is produced, also known as spin-orbit coupling. The spin-spin coupling can also induce precession, but it is expected to be much smaller than the geodesic precession at large
separations\citep{gerosa2015multi}. From the Eq. (2.4a) in \citep{kidder1995PhRvD52821K}, the last two terms are the spin-spin coupling, due to the magnitude of orbital angular momentum is usually several times larger than the spin magnitude (always less than 1 for black holes), and also considering the coefficient of the spin-orbit term,  the spin-spin one will several times less than the spin-orbit coupling. 
For simplicity in the present work, we only consider the dominant effect (spin-orbit coupling), and this will not affect the result. For a binary of total mass $M=m_1+m_2$, relative separation $r=|\mathbf{r}|$ and spins $\mathbf{S}_1=\chi_1 m_{1}^2\hat{\mathbf{S}}_1$ and $\mathbf{S}_2=\chi_1 m_{2}^2\hat{\mathbf{S}}_2$, where $\hat{\mathbf{S}}_{1,2}$ are unit vectors and dimensionless spin parameter $0<\chi_{1,2}<1$, the precession equations of the two binary components are given by: \citep{Buonanno2003PhRvD}

\label{ds1}
\begin{align}
    \dot{\mathbf{S}}_{1} &=\frac{1}{r^{3}}\left[\frac{4 m_{1}+3 m_{2}}{2 m_{1}}\left(\frac{G\mu} {c^{2}} (GM)^{1 / 2} r^{1 / 2}\right) \hat{\mathbf{L}}\right] \times \mathbf{S}_{1} \,,
\end{align}
just replace 1 with 2, one can get the equation for $\mathbf{S}_2$, where $\mu=m_1m_2/M$ is the reduced mass and $\hat{\mathbf{L}}\propto\mathbf{r}\times\dot{\mathbf{r}}$ gives the direction of the orbital angular momentum. From Eq. (\ref{ds1}), we get the precession rates is

\begin{align}
    \Omega_{S_1} \approx 2 \times 10^5 F_1(q)  \frac{M_\odot}{M}\tilde{r}^{-5/2}, \,\: \Omega_{S_2} \approx 2 \times 10^5 F_2(q)  \frac{M_\odot}{M}\tilde{r}^{-5/2} \,,
\end{align}

where $\tilde{r}=r/(GM/c^2)$, $q$ is the mass-ratio defined by $q=m_2/m_1$, and 
\begin{align}
F_1(q) = \frac{q(4+3q)}{2(1+q)^2} \,,\:F_2(q) = \frac{q(4q+3)}{2(1+q)^2} \,.
\end{align}
It is clear that when $q \to 0$, the precession of $\mathbf{S}_1$ disappears and the one of $\mathbf{S}_2$ just returns to the well-known geodesic precession. For equal-mass binary $F_1 = F_2 = 7/8$.

Considering that the Keplerian orbital frequency has a relation $\omega_{\rm orb} \approx 2 \times 10^5 (M_\odot/M)\tilde{r}^{-3/2}$, we can get the relation between the frequencies of inspiral's and precession's GWs. For the equal-mass case, if the mass of each black hole is 10 (100) $M_\odot$, when the objects inspiral from $10$ to $3M$, then the precession frequency of each spin is from 9 (0.9) Hz to 179 (17.9) Hz, which is in the sensitive band of Advanced LIGO, Virgo and KAGRA detectors.

We can then calculate the waveform emitted by precessing black holes. The waveform of a precessing rigid body can be written as \citep{maggiore2007gravitational}
\begin{align}
\label{ht}
    h_+ = A_{+,1}\cos{\Omega t} + A_{+,2}\cos{2\Omega t} \,,\:
    h_\times = A_{\times,1}\cos{\Omega t} + A_{\times,2}\cos{2\Omega t} \,,
\end{align}
where 
\begin{align}
\label{aa}
 A_{+, 1}&=h_{0}^{\prime} \sin 2 \alpha \sin \iota \cos \iota\,, \:A_{+, 2}=2 h_{0}^{\prime} \sin ^{2} \alpha\left(1+\cos ^{2} \iota\right), \\
 A_{\times, 1}&=h_{0}^{\prime} \sin 2 \alpha \sin \iota\,, \: A_{\times, 2}=4 h_{0}^{\prime} \sin ^{2} \alpha \cos \iota \,,  \label{atimes} 
\end{align}
where $\iota$ is the angle between the line-of-sight and the initial(non-precessing) spin direction, $\alpha$ is the spin rotation angle and
\begin{align}
    h_{0}^{\prime}=-\frac{G}{c^{4}} \frac{\left(I_{3}-I_{1}\right) \Omega^{2}}{D} \,. \label{h0I}
\end{align}
We notice that the waveform has two frequencies, one is equal to the precession frequency $\Omega$, and the other one is the double of $\Omega$. For sufficiently strong precessing effect $\alpha\geq\pi/4$, by Eqs.~(\ref{aa},\ref{atimes}), the $2\Omega$ component is several times larger than the $\Omega$ one, therefore we mainly focus the dominant part.   

The amplitude of the waveform is determined by the difference in the moments of inertia, $I_3$ is related to the rotation about the direction of the total angular momentum and $I_1$ is related to the principal axis perpendicular to this direction. The principal moment of inertia can be obtained from the following equation~\citep{hartle2003gravity}
\begin{align}
    J =\frac{G m}{c^2}a=2 \frac{G^2m^3}{c^4}(1+\sqrt{1-\chi^2})\Omega_{\rm H},
\end{align}
where $m$ is the mass, $a$ is the Kerr parameter, $J$ is the total angular momentum of the Kerr black hole and $\Omega_{\rm H}$ is the rotating frequency of horizon. One then can define the main inertia moment by $J = I \Omega$ as
\begin{align}
    I_3 =  2 \frac{G^2m^3}{c^4}(1+\sqrt{1-\chi^2})\,, \label{I3}
\end{align}
which depends on the spin $S$ of the black hole. Therefore, $I_1 - I_3$ can be taken as $\delta G^2m^3/c^4$, where $\delta$ is a function determined by the BH mass and spin in GR or by several other parameters in alternative gravity theories. 
Finally, the expression of the GW strain amplitude given by Eq.(\ref{h0I}) becomes:
\begin{align}
    h_{0i}^{\prime}=-\frac{G^3}{c^{8}} \frac{\delta m_i^3 \Omega_{S_i}^{2}}{D} \approx 4.8 \times 10^{-23} \left(\frac{1 {\rm Gpc}}{D}\right)\left(\frac{m_i}{m_\odot}\right)\left(\frac{m_i}{M}\right)^2 \delta F_i(q)\tilde{r}^{-5} \,, \label{h0}
\end{align}
where $i=\{1,2\}$. From the above equation, the GW strain amplitude of the signal emitted by precessing black holes increases very rapidly when r decreases. For $r$ small enough, we will see that this signal can be detected.

To do so we need to derive the frequency-domain waveform of the GWs from spin precession. To the leading order, the orbital evolution due to the GWs radiation for a circular binary is  
\begin{align}
 - \frac{G \mu M}{2 r^2}\dot{r}=\frac{32 G \mu^2 r^4 }{5 c^5} \left(\frac{G M}{r^3}\right)^3\,, \label{energybalance}
\end{align}
here and after, we omit the subscript $i$ of $m$ and $F$. From Eq.(\ref{ds1}) we can get the relation between the precession frequency and the orbital radius
\begin{align}
   r=c^{-4/5} G^{3/5} M^{3/5} F^{2/5} \Omega_S ^{-2/5} \,.\label{rtomega}\end{align}
Using the above expression of $r$ and the definition of the angular frequency $f_p\equiv\Omega_S/2\pi$, the energy balance equation\eqref{energybalance} can be rewritten in a form of a differential equation of $f_p$ with respect to time $t$, as
\be
 \dot{f}_p=\frac{2^{33/5} \pi ^{8/5} G^{3/5} \mu}{c^{9/5} M^{2/5} F^{8/5}}f_p^{13/5} \,.\label{dotf}
\ee
while the equation for inspiral orbital frequency is $\dot{f}\propto f^{11/3}$, it will induce the waveform of precession's GWs  have a different power law with respect to frequency. In terms of $\tau = t_c-t$, where $t_c$ is the coalescence time, the solution of Eq.\eqref{dotf} reads
\begin{align}
f_p(\tau)=\frac{5^{5/8} c^{9/8} F_1 M^{1/4}}{64 \pi  G^{3/8} \mu^{5/8}} \tau ^{-5/8}\,.\label{ftot}
\end{align}
Using Eq.\eqref{ftot} and the phase of precession's GWs $\Phi(t)=\int_{t_0}^t\!\Omega_S d t^{'}$, we find (here $d \tau=-d t$ and we use the signature $(-,+,+,+)$ for the metric)
\begin{align}
&\Phi(\tau)=-\frac{5^{5/8} c^{9/8} F_1 M^{1/4} }{12 G^{3/8} \mu^{5/8}}\tau ^{3/8}+\Phi_c\,\label{phit},\\
&\ddot\Phi(\tau)=\frac{ 5^{13/8} c^{9/8} F_1 M^{1/4}}{256 G^{3/8} \mu^{5/8}}\tau ^{-13/8}\,\label{ddotphi}.
\end{align}
Equivalently, using the stationary point condition $2\pi f_p=\dot\Phi(t_*)$ and $\tau_*=t_c-t_*$, we can rewrite \eqref{ftot} as
\begin{align}
&\tau_*(f_p)=\frac{5 c^{9/5} M^{2/5} F^{8/5}}{2^{48/5} \pi ^{8/5} G^{3/5} \mu}f_p^{-8/5}\,\label{ttof}.
\end{align}
We can, now, compute the Fourier transform of the  $\Omega$ component of $h_+$ in Eq.\eqref{ht} using the stationary phase method~\citep{cutler1994gravitational}. \comment{Since $d \ln A_{+,1}/dt\ll \dot\Phi(t)$ and $\ddot\Phi(t)\ll(\dot\Phi(t))^2$, the stationary phase approximation provides the following estimate of the Fourier transformation}
Finally, the frequency domain waveform from the precession is given by:
\begin{align}
\tilde{h}_{+,1}=-\frac{\pi ^{6/5} \delta  f^{7/10} G^{27/10} m^3 M^{1/5} F_1^{4/5}}{2^{33/10} c^{71/10} \mu^{1/2} D} e^{i \psi_{+,1}} \sin 2 \alpha  \sin 2 \iota \,,\label{hpf}
\end{align}
where
\begin{align}
\psi_{+,1}=2 \pi  f \left(t_c+\frac{D}{c} \right)-\Phi_c-\frac{\pi }{4}+\frac{25 c^{9/5} M^{2/5} F_1^{8/5}}{768 (2 \pi )^{3/5} \mu(f G)^{3/5}}\,.
\end{align}
By repeating the same procedure for $\tilde{h}_{+,2}$ we can derive the dominant $2\Omega$ component ($f_2=2f$) waveform
\comment{\begin{align}
\tilde{h}_{+}=-\frac{\pi ^{6/5} \delta  f^{7/10} G^{27/10} m^3 M^{1/5} F_1^{4/5} }{2^{33/10} c^{71/10} \mu^{1/2} D}\left(e^{i \psi_{+,1}} \sin 2 \alpha  \sin 2 \iota +16 e^{i \psi_{+,2}} \sin ^2\alpha \cos ^4\frac{\iota }{2}\right)\,,
\end{align}
where
\begin{align}
\psi_{+,2}=4 \pi  f \left(t_c+\frac{D}{c} \right)-\Phi_c-\frac{\pi }{4}+\frac{25 c^{9/5} M^{2/5} F_1^{8/5}}{192 (2 \pi )^{3/5} \mu(f G)^{3/5}}\,.
\end{align}
\begin{align}
f_2=F_1 \left(\frac{\pi G M}{c^3}\right)^{2/3}  f_{GW}^{5/3}\,.
\end{align}}
\begin{align}
\tilde{h}_{+,2}=-\frac{\pi ^{6/5} \delta  f^{7/10}_2 G^{27/10} m^3  M^{1/5} F_1^{4/5} }{ c^{71/10} \mu^{1/2} D} e^{i \psi_{+,2}} \sin ^2\alpha \cos ^4\frac{\iota }{2}\,,
\label{hpf2}
\end{align}
where
\begin{align}
\psi_{+,2}=2\pi  f_2 \left(t_c+\frac{D}{c} \right)-\Phi_c-\frac{\pi }{4}+\frac{25 c^{9/5} M^{2/5} F_1^{8/5}}{192 \mu(\pi f_2 G)^{3/5}}\,.
\end{align}

In the following calculation we will only consider this dominant part which radiates at $f_2$, so $\tilde{h}_+=\tilde{h}_{+,2}$. Note that this is only the gravitational wave of the precession of $\mathbf{S}_1$, similarly one can also calculate the precession's waveform of the component black hole. For equal-mass binary black holes with the same spin magnitude, the total GW strain from spin precession is exactly twice that of the $\mathbf{S}_1$ component. From Eqs. (\ref{hpf},~\ref{hpf2}), we can clearly see that the frequency-domain strain is proportional to $f^{7/10}$, which distinguishes from the inspiral one $f^{-7/6}$. 

\begin{figure}
\centering
{
\includegraphics[width=\columnwidth]{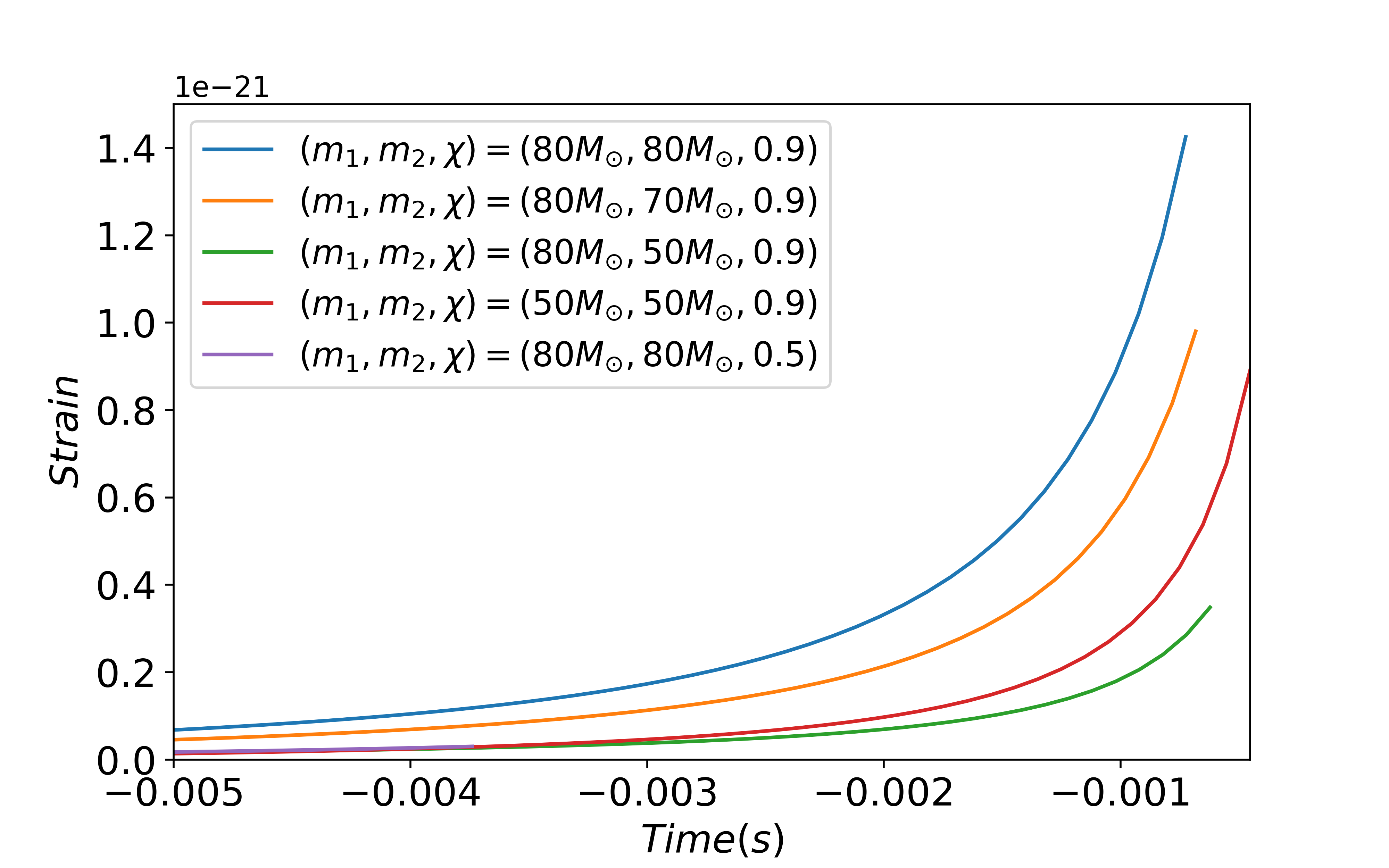}
}
\caption{\emph{Five burst precession's waveforms.} All the waveforms are cut off at the light-ring.}
\label{fig:waveform}
\end{figure}

Fig.~\ref{fig:waveform} shows five time-domain precession's waveforms that are cut off at the light ring. This very short signal, called burst, is very different from the inspiral signal. Furthermore, we can observe that
binaries of nearly equal mass have a stronger precession's signal than unequal cases. Faster rotating and more massive binaries can produce stronger signals.   

\section{Results and detectability}

\begin{table}
\begin{center}
\caption{SNR of some precession's GW signals detected by Advanced LIGO and ET}
\begin{tabular}{|c|c|c|c|c|c|}
\hline$(m_1,m_2,\chi)$&$80,80,0.9$&$80,70,0.9$&$80,50,0.9$&$50,50,0.9$&$80,80,0.5$\\ \hline
aLIGO&$17.4$&$11.1$&$3.4$&$9.4$&$0.6$\\ \hline
ET&$167.3$&$106.7$&$32.3$&$98.1$&$6.1$\\ \hline
\end{tabular}
\label{SNR}
\end{center}
\end{table}

To obtain the magnitude of the precession's waveforms, we need the value of $\delta$ which is determined by the difference between $I_1$ and $I_3$. The main inertia $I_3$ is well known for the Kerr black hole. In another way, one can get the inertial moment using the radii of gyration of the event horizon. The rotation-offset radii of horizon is
\begin{align}
    \sigma_+ = \frac{Gm}{c^2} [(1+\sqrt{1-\chi^2})^2+\chi^2\cos^2{\theta}]^{1/2} \,,
\end{align}
then the gyration radius is $(2Gm/c^2 \sigma_+)^{1/2}$. Taking $\theta = \pi/2$ we get the main gyration radius
\begin{align}
    R_z = [2\frac{G^2m^2}{c^4}(1+\sqrt{1-\chi^2})]^{1/2} \,,
\end{align}
then the main inertial moment is $m R_z^2$ and just equals to Eq. (\ref{I3}). Due to the axisymmetry, $I_1 = I_2$ and thus, taking $\theta = 0$ we have
\begin{align}
    I_1 = 2\frac{G^2m^3}{c^4}[(1+\sqrt{1-\chi^2})^2+\chi^2]^{1/2} \,.
\end{align}
Therefore, $I_1 - I_3$ can be taken as $\delta G^2m^3/c^4$ with $\delta = 2[(2\tilde{r}_+)^{1/2}-\tilde{r}_+]$ with $\tilde{r}_+ \equiv r_+/(Gm/c^2)$. If we adopt the above calculation, then for $0 \leq \chi\leq 1$, we get $0 \leq \delta \leq 2(\sqrt{2}-1)$. For $\chi = 0.9$, $\delta \approx 0.52$.   

We will now discuss the detection potential of such signals. To simplify the calculation, we assume that $\chi_1=\chi_2=\chi$. 
Using Eq.\eqref{ds1} we obtain the peak frequency of GWs radiated from spin precession which cut off at the light ring \citep{bardeen1972rotating} (the precession's GWs after merger are not considered in our present analysis.)
\begin{align}
f_{\rm peak}=\frac{c^3}{2\pi G M}F\left(\frac{c^2}{G M}r_{\rm LR}\right)^{-5/2}\,.
\end{align}
It is clear that the peak frequency increases with $|\chi|, F$ and decreases with $M$, and $F$ positively correlated with mass ratio $q$. In addition, from Eq.\eqref{hpf2} the strain of precession's GWs will be largely reduced for relatively small values of mass ratio. Therefore, the SNR will be larger when the spin becomes extreme and the binary black holes have nearly equal mass. We then calculate the SNR of the precession's signal measured by Advanced LIGO and ET(Table~\ref{SNR}). From the events detected by Advanced LIGO and Virgo, we choose a comparatively reliable parameter space: mass $50-80M_\odot$ and spin $0.5 - 0.9$(such as two extreme spin BH
binary events reported in the GWTC-2 catalog, GW190517{\_}055101 has $\chi_{1,2}>0.8$ with 77\% credibility, and GW190521 with 58\% credibility\citep{abbott2021gwtc}). Moreover, since the SNR is inversely proportional to the distance from the source, we set a  fiduciary value of 100 Mpc in our analysis. We notice that for Advanced LIGO, the signal is detectable (SNR$\geq 10$) only if the masses of the black holes are nearly equal ($\geq 50M_\odot$ each) with an extreme spin ($\chi\geq 0.9$) and a luminosity distance around 100 Mpc.
On the other hand, we found that the SNR in ET is almost 10 times higher than that in AdvancedLIGO.

\begin{figure}
\centering
\includegraphics[width=\columnwidth]{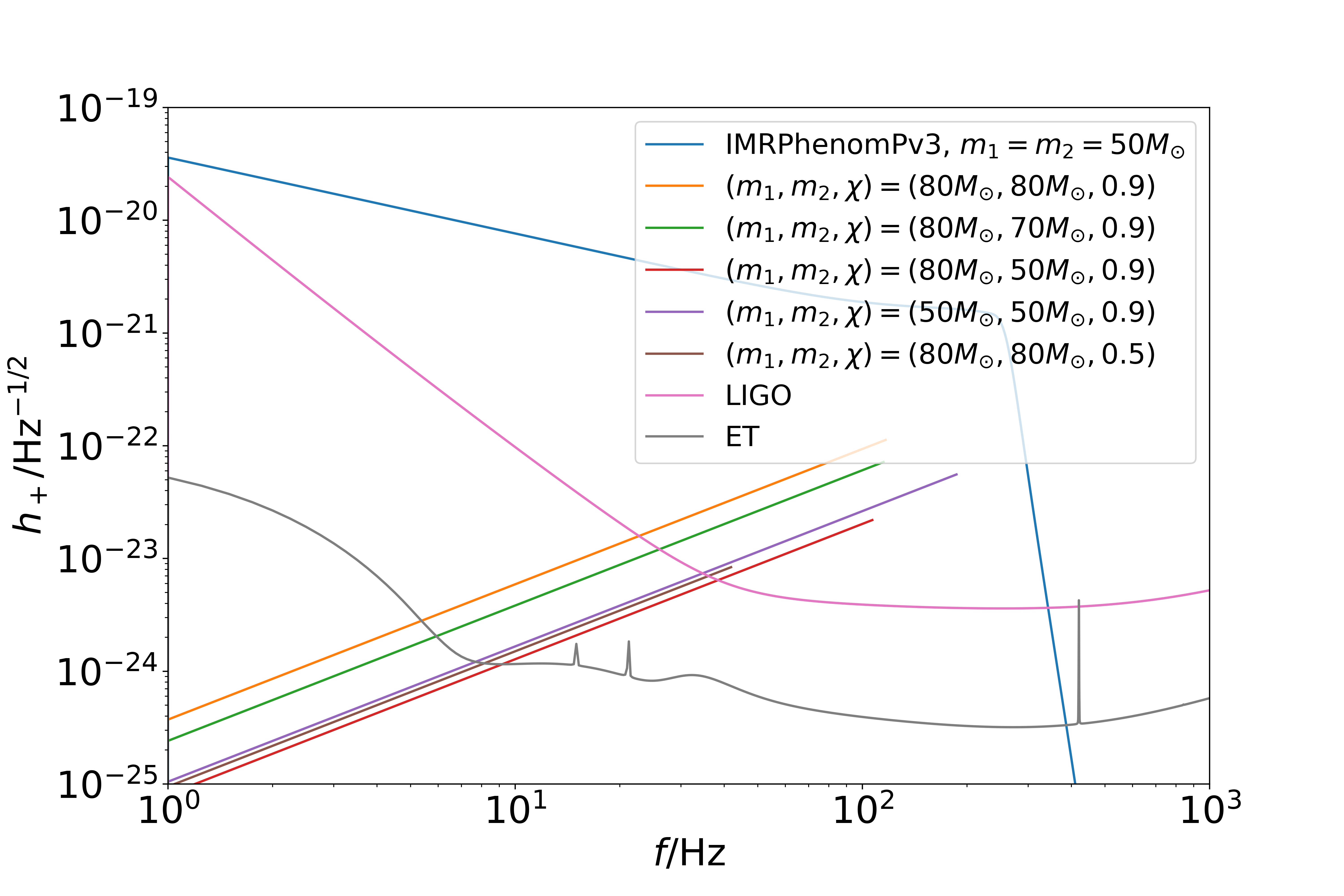}
\caption{\emph{PSD of five precession's signals and one inspiral-merger-ringdown signal}. Three equal-mass(50 and 80$M_\odot$) high spin($\chi=0.9 \& 0.5$) binaries are all set with parameters $\alpha=\pi/4$, distance is 100 Mpc and face-on($\iota=0$) to the observer, and an IMR signal($m_1\!=\!m_2\!=\!50M_\odot$) for comparison are shown. The other two imaginary precession's signals are unequal-mass($m_1=80M_\odot,m_2=70M_\odot \& m_1=80M_\odot,m_2=50M_\odot$) extrem spin($\chi=0.9$) binaries.}
\label{strain}
\end{figure}

Based on our estimation, the precessive GW signal with appropriate parameters should be strong enough to be measurable by the Einstein telescope. Therefore, in order to determine the three main parameters $(m_1, m_2, \chi)$ that contribute to the precessive GW signal, we assume that the distance of 
the source is 100 Mpc and that it is facing the observer($\iota=\pi/2$). Fig.~\ref{strain}  shows the power spectral density (PSD) of five precession's signals and one IMR (inspiral-merger-ringdown) signal. Obviously, the precessive GW signal is much weaker than the IMR signal for the same source. However, some sources could be detected even by AdvancedLIGO if the mass of each black hole is as large 
as GW190521 and as close as a few hundred Mpc with high spin and precession(see Table.~\ref{SNR}). For third generation detectors, precessive GW signals could be more easily detected by ET. Furthermore, due the spin-spin effect just a few times smaller than the one of spin-orbit, in ET era, it is also possible to detect the spin-spin effect. This will be discussed in a future work.

After passing, the light ring, two BHs merge and a final black hole is formed. During this process, we believe that the rotation axis of the dynamical BH continues to precess for some time. The radiation from the precessive GWs will reduce the precession itself to zero eventually. In this Letter, we do not consider the waveform in this final stage, but it should be  very interesting in a future work.


\section{ Conclusions}

The GWs directly from precessing black hole in binary system usually are too weak to be detected by Advanced LIGO. However, if this type of source is within a few hundred Mpc, and if it is the merger of two 
relatively massive black holes ($m_{1,2} \gtrsim 50 M_\odot$) with large spin precession and a comparable mass-ratio, then the precession's signals should be detectable by current ground-based detectors. For example, as shown in Table~\ref{SNR}, if the source is coalescence of equal-mass black holes($m1=m2=80M_\odot$) with extreme spin($\chi=0.9$), we obtain the SNR is 17.4 for Advanced LIGO. For the third generation detectors like ET, this kind of GWs would be detected more easily. Further more, due the significantly different chirp property, this kind of GWs can be easily  distinguished  from the main inspiral signal in the data.

Considering that the nature of this kind of precessing waveforms comes from the inertia moments of spinning black hole, then the inertia and quadrupole of BH will directly decide the strength of waveforms. However, the degeneration still exists, the spin magnitude, precession frequency, angle and so on also influence the waveform strain. Fortunately the degeneration could be broken, if we combine the inspiral signals. Then one can measure the $I_3 - I_1$ which is determined by the structure of the rotating black holes from the precession's signals. 

In GR, the astrophysical BH should be described by the Kerr solution, which only has two parameters: mass and angular momentum. The quadrupole moment will be determined by $M, ~\chi$ uniquely. In principle, the quadrupole decides the form of precession's waveforms. However, alternative gravity theories may predict hairy black holes, i.e., at least one more parameter will be introduced to give quadrupoles. From precession's GWs, in principle one can directly measure the quadrupole, in other words,  the no-hair theorem will be tested. This means that this kind of signals could be a unique tool to reveal the instinct structure of black holes.

\section*{Acknowledgements}
This work is supported by The National Key R\&D Program
of China (No. 2021YFC2203002), NSFC No. 11773059 and No. 12173071,  and the Strategic Priority Research Program of the CAS under Grants No. XDA15021102. W. H. is supported by CAS Project for Young Scientists in Basic Research YSBR-006. We thank Prof. Y. Chen for his useful comments. 

\section*{DATA AVAILABILITY}
The data underlying this article will be shared on reasonable request to the corresponding author Wen-Biao Han.
%





\bsp	
\label{lastpage}

\end{document}